\documentclass[conference]{IEEEtran}
\IEEEoverridecommandlockouts
\usepackage{cite}
\usepackage{amsmath,amssymb,amsfonts}
\usepackage{algorithmic,multicol}
\usepackage{graphicx}
\usepackage{multirow}
\usepackage[table,xcdraw]{xcolor}
\usepackage{textcomp}
\usepackage{xcolor}
\usepackage[ruled,vlined,linesnumbered]{algorithm2e}
\def\BibTeX{{\rm B\kern-.05em{\sc i\kern-.025em b}\kern-.08em
    T\kern-.1667em\lower.7ex\hbox{E}\kern-.125emX}}
\begin{document}

\title{MAVSec: Securing the MAVLink Protocol for Ardupilot/PX4 Unmanned Aerial Systems\\
{\footnotesize \textsuperscript{}}
\thanks{The paper is accepted in the International Wireless Communications \& Mobile Computing Conference (IWCMC) in Morocco, June 2019.}
}

\author{
\IEEEauthorblockN{ Azza Allouch}
\IEEEauthorblockA{\textit{Faculty of Sciences of Tunis (FST)} \\
\textit{University of El Manar, Tunis, Tunisia}\\
\textit{LISI Laboratory, (INSAT)}\\
azza.allouch@coins-lab.org}
\and
\IEEEauthorblockN{Omar Cheikhrouhou }
\IEEEauthorblockA{\textit{College of CIT, Taif University}\\
\textit{Taif, Saudi Arabia} \\
\textit{Computer and Embedded Systems Laboratory}\\
 \textit{University of Sfax, Sfax, Tunisia}\\
o.cheikhrouhou@tu.edu.sa}
\and

\IEEEauthorblockN{Anis Koub\^{a}a}
\IEEEauthorblockA{\textit{Prince Sultan University, Saudi Arabia} \\
\textit{CISTER/INESC-TEC, ISEP, Portugal}\\
\textit{Gaitech Robotics, China}\\
 akoubaa@psu.edu.sa}

\and
\IEEEauthorblockN{ Mohamed Khalgui}
\IEEEauthorblockA{\textit{School of Electrical and Information Engineering} \\
 \textit{Jinan University, China} \\
\textit{LISI Laboratory, (INSAT)} \\
\textit{ University of Carthage, Tunis, Tunisia}\\
khalgui.mohamed@gmail.com}

\and
\IEEEauthorblockN{ Tarek Abbes}
\IEEEauthorblockA{\textit{Digital Security Research Unit, ENETCOM} \\
\textit{University of Sfax, Sfax, Tunisia}\\
tarek.abbes@enetcom.usf.tn}

}

\maketitle

\begin{abstract}
The MAVLink is a lightweight communication protocol between Unmanned Aerial Vehicles (UAVs) and ground control stations (GCSs). It defines a set of bi-directional messages exchanged between a UAV (aka drone) and a ground station. The messages carry out information about the UAV's states and control commands sent from the ground station. However, the MAVLink protocol is not secure and has several vulnerabilities to different attacks that result in critical threats and safety concerns. Very few studies provided solutions to this problem. In this paper, we discuss the security vulnerabilities of the MAVLink protocol and propose MAVSec, a security-integrated mechanism for MAVLink that leverages the use of encryption algorithms to ensure the protection of exchanged MAVLink messages between UAVs and GCSs. To validate MAVSec, we implemented it in Ardupilot and evaluated the performance of different encryption algorithms (i.e. AES-CBC, AES-CTR, RC4 and ChaCha20) in terms of memory usage and CPU consumption. The experimental results show that ChaCha20 has a better performance and is more efficient than other encryption algorithms. Integrating ChaCha20 into MAVLink can guarantee its messages confidentiality, without affecting its performance, while occupying less memory and CPU consumption, thus, preserving memory and saving the battery for the resource-constrained drone.
\end{abstract}


\section{introduction}
Autonomous Unmanned Aerial Vehicles (UAVs) are an emerging technology that has attracted several applications such as smart cities, \textcolor{black}{border surveillance }, traffic monitoring \cite{8658300}, security, natural disaster monitoring, \textcolor{black}{real-time object tracking\cite{koubaa2018dronetrack}} and transport \cite{pajares2015overview,hayat2016survey}.

These flying vehicles are controlled either remotely from a Ground Control Station (GCS) or autonomously by a pre-programmed mission. When controlled remotely, the communication between the UAV and the GCS is established through a communication protocol. The Micro Air Vehicle link (MAVLink) \cite{li2013communication} is one of the most widely used protocols for communication between UAVs and GCSs. MAVLink was developed to be a flexible, lightweight, open source communication protocol specifically used for the bidirectional data exchange between the autopilot and the GCS. The GCS sends commands and controls to the drone, while, the latter sends telemetry and status information data \cite{veena2014towards} to the GCS. MAVLink is also used to connect drones over the Internet \cite{ICARSC2017, koubaa2018dronetrack, koubaa2019dronemap,Qureshi:2016}.

MAVLink is used by several autopilot systems including Ardupilot \cite{ardupilot} and PX4 \cite{px4}. Ardupilot and PX4 are the leading open source autopilot systems designed to control any type of unmanned vehicles, including fixed-wing aircraft, and various rotary-wing platforms, namely single, tri, quad, hexa, octa copters and even submarines \cite{ardupilot}. PX4 also offers similar capabilities for UAVs and can be extended to underwater systems. 

Despite, the wide use of the MAVLink protocol it presents vulnerabilities and is prone to several attacks including spoofing, message forging and denial of service (DoS) as proven in\textcolor{black}{\cite{koubaa2019dronemap}} and \cite{kwon2018empirical}. These vulnerabilities are mainly due to the fact that the protocol does not implement any security mechanism and does not adopt any encryption algorithm. Therefore, the GCS communicates with the UAV over an unencrypted channel, thus subject to several types of attacks.

In literature, a few studies have discussed possible security solutions for the MAVLink protocol. In \cite{rajatha2015authentication},\cite{Hamsavahini}, the authors used the Caesar cipher cryptography for data encryption of MAVLink messages between the ground station and the Micro Aerial Vehicles (MAV). They showed that a secret key was clearly sent from the GCS to the drone, during the establishment phase. An intruder, who may eavesdrop the communication, could easily detect the key, and thus can break all the security system. Moreover, the Caesar encryption algorithms used in these works are known to be insecure and vulnerable to crypt-analysis. 
In \cite{butcher2013securing}, the authors addressed the MAVLink protocol security against passive attacks such as eavesdropping and interception, and implemented the RC5 encryption algorithm to encrypt the MAVLink messages. However, the proposed method lacks an authentication mechanism that protects communication from active attacks such as forging the message, identity spoofing, etc. It is worth noting that a secure version of MAVLink is currently being discussed by the protocol developers, but has not yet been developed \cite{restt}.
 
In this paper, we suggest improving the MAVLink protocol security in order to protect the communication between drones and GCSs. This allows to mitigate malicious attacks. The proposed method involves the implementation of several encryption algorithms, namely, RC4, AES-CBC, AES-CTR and ChaCha20 to ensure the confidentiality of the exchanged messages between UAV and GCS. We also evaluate their performances in terms of memory usage and CPU consumption. 


In summary, the four main contributions of this paper are as follows:

\begin{itemize}
    \item First, we identify the MAVLink protocol security threats.
\end{itemize}
\begin{itemize}
    \item Second, we propose MAVSec, a MAVLink enhanced version with cryptographic mechanisms to ensure confidentiality of the exchanged messages between UAVs and GCSs.
\end{itemize}
\begin{itemize}
    \item Third, we implement the security mechanisms in Ardupilot using the MAVLink protocol to demonstrate the feasibility of the proposed solutions.
\end{itemize}
\begin{itemize}
    \item Fourth, we prove the effectiveness through performance evaluation of our proposal.
\end{itemize}
The remainder of this paper is organized as follows. Section II discusses the related works. Section III describes the MAVLink protocol. Section IV presents security threats against the MAVLink protocol. Section V describes the proposed cryptographic mechanisms to secure the MAVLink protocol. Detailed simulation and experimental results are presented and discussed in section VI. Finally, Section VII provides some concluding remarks.


\section{Related works}
Ensuring security has become increasingly necessary and important for the wide adoption of UAV systems. The existing security methods proposed for UAV systems can be classified into hardware \cite{shoufan2015secure,yoon2017security} and software approaches \cite{rajatha2015authentication,butcher2013securing,marty2013vulnerability}.
 
In \cite{shoufan2015secure}, a hardware-based implementation of the AES protocol was proposed to secure the communication between a  ground control station (GCS) and the drone. An FPGA module, connected to the drone embeds the cryptographic solution: AES-CBC-MAC, was used to encrypt and authenticate both commands and payload data transmitted between the drone and the GCS. However, the hardware solution affects negatively the system performance and power consumption due to the extra hardware weight.
 
In \cite{yoon2017security}, the authors proposed the idea of an additional encrypted communication channel to enhance the security of data in UAVs through Raspberry Pi. This channel was designed to regain control of the UAV in case it was target to attack. However, this hardware solution displays time delay between GCS and Raspberry Pi and increases the CPU usage on  Raspberry Pi. The experimental setup is not applied to real drones' communication.
 
In the context of software based solutions, the authors in \cite{rajatha2015authentication} proposed a methodology for data encryption and authentication of MAVLink messages between the ground station and the UAV using Caesar cipher cryptography. However, its main drawback is that it can be easily broken. Moreover, the results have not been explicitly stated. In our paper,however, we implement robust cryptographic methods and clearly present our simulation results.

An encryption mechanism RC5 was used in \cite{butcher2013securing} to secure the MAVLink communication protocol. This study only provided a description of the protocol without any specific details of testing and performance analysis. In our paper, we identify and evaluate the performance of four cryptographic methods used to secure the MAVLink protocol.

In \cite{marty2013vulnerability}, Marty presented a vulnerability analysis of the MAVLink protocol, suggesting some cryptographic algorithms to secure the MAVLink protocol and provide a methodology to evaluate the cost of securing the MAVLink protocol. However, in their work, the cryptographic techniques were not implemented and the feasibility of the approach was not shown. In our paper, we implement the encryption mechanisms into the source code of the Ardupilot to enable a secure communication using MAVLink protocol along with a performance evaluation.

\section{MAVlink System Architecture}
MAVLink is an open source, lightweight and header-only protocol mostly used for bidirectional communications between GCSs and UAVs. MAVLink 1.0 was first released in early 2009 by Lorenz Meier under LGPL license \cite{qground}. \textcolor{black}{The MAVLink 2.0 protocol \cite{mavlinkv2} was released in early 2017 and is the current recommended version. It is backward compatible with the MAVLink 1.0 version and includes several improvements over the MAVLink 1.0 version.} 

\begin{figure}[htb!]
	\centering
	\includegraphics[width=0.5\textwidth]{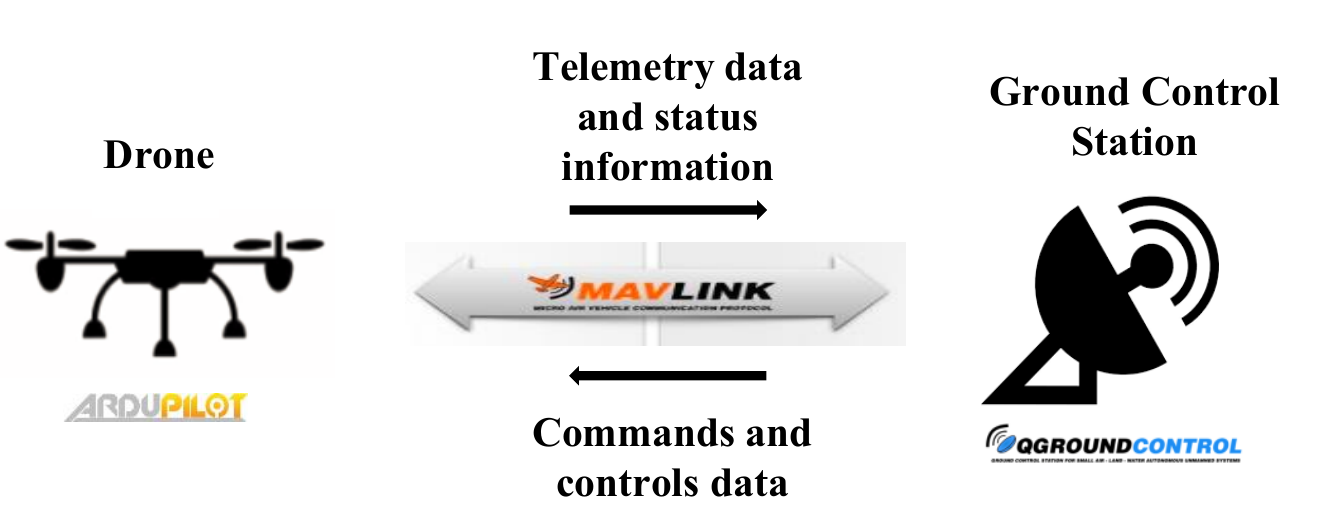}
	\caption{Communication link between UAV and GCS.}
	\label{fig.1}
\end{figure}

\begin{figure*}[htb]
\centering
\includegraphics[width=0.9\textwidth]{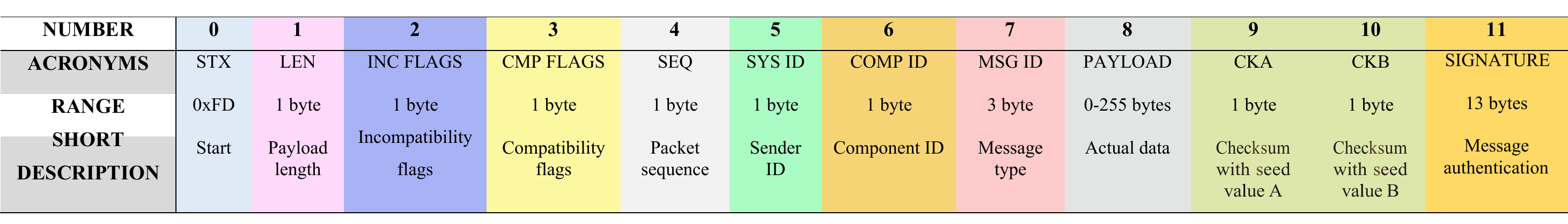}
\caption{MAVLink 2.0 packet structure.}
\label{MAVLinkv2}
\end{figure*}

The MAVLink messages are of two types: (\textit{i.}) commands and control messages transmitted from the GCS to the UAV, and, (\textit{ii.}) state information messages (e.g., position, heartbeat, and system status information) sent from the UAV to the GCS, as depicted in Fig.\ref{fig.1}. Since the MAVLink protocol is used for real-time communication; it is designed to be a lightweight protocol. Figure \ref{MAVLinkv2} shows the structure of the MAVLink 2.0 packet.

All MAVLink messages contain a header appended to each data payload of the message. The header contains information about the message while the payload contains the data carried out by the message. The checksum is intended to verify the integrity of the message, that should not be altered during its transmission.

\textcolor{black}{MAVLink protocol is a variable size protocol. 
The minimum packet length of a MAVLink message is 11 bytes (\textit{STX}, \textit{LEN}, \textit{INC FLAGS}, \textit{CMP FLAGS}, \textit{SEQ}, \textit{SYS ID}, \textit{COMP ID}, \textit{MSG ID}, \textit{CKA} and \textit{CKB}) and the maximum packet length, with full \textit{payload} and \textit{signature}, is 297 bytes. The payload size is variable, its length depends on the parameters which are sent or received during the communication. The signature field allows the authentication of the message and verifies that it originates from a trusted source.}

MAVLink message types are identified by the ID field on the packet, and the payload contains the appropriate data. Several control and state messages are defined in the MAVLink protocol. The most crucial message in MAVLink is the heartbeat message. Initially, a drone should send the HEARTBEAT message periodically (generally every second) to the ground station to provide feedback of their status (to indicate that the drone is active and still connected). This is a mandatory message.


\section{Security Threats}
With the increasing use of UAVs in military and civilian applications, they are carrying sensitive and secure information that can be sniffed by attackers. In fact, the MAVLink protocol does not provide any kind of security and can be hacked quite easily. There is no confidentiality, nor authentication mechanism. The GCS communicates with drones over an unauthenticated and unencrypted channel. Anyone with an appropriate transmitter can communicate with the drone and inject commands into an existing session, and thus can easily impersonate any drone. Also, MAVLink message streams can be easily intercepted and eavesdropped  by hackers because they are sent with no encryption.  

According to \cite{marty2013vulnerability}, the MAVLink protocol is vulnerable to attacks and does not provide the CIA (Confidentiality, Integrity and Availability) security services. Thus, the MAVLink protocol could be exposed to different attacks such as Interception (Attacks against the system’s confidentiality), Modification (Attacks against the system’s integrity), Interruption (Attacks against the system’s availability).

Interception can be achieved by eavesdropping on channels. The message contents are read by unauthorized users. Since the MAVLink communication protocol is not always secured, an intruder is able to intercept information about commands sent to the UAV from GCS and steal other data sent in the opposite direction. Authentication and encryption should be used on the link to mitigate this risk and guarantee the confidentiality and integrity of the exchanged data. 
 
Modification means tampering with an original message. MAVLink protocol does not ensure integrity which might allow an attacker to effectively hijack the UAV from its GCS. If there is no integrity protection mechanisms, malicious attacks on the network or wireless channel interference may cause information modification, and thus become invalid.
 
Interruption means that a message from/to a particular service is blocked. An adversary disables the reception of MAVLink control signals from the ground control by the drone. The communication between the drone and the GCS is blocked, and the aircraft will go into a lost link state. Implementing strong authentication mechanisms can help mitigate the risk of unavailability.
Various security threats against the MAVLink protocol, with the corresponding mitigation techniques are listed in Table \ref{tab:threats}.

\begin{table*}[]
\caption{\label{tab:threats} Security threats on MAVLink and countermeasures.}
\centering
\begin{tabular}{|c|c|c|}
\rowcolor[HTML]{C0C0C0} 
\hline
\textbf{Security objective} & \textbf{Threats} & \textbf{Mitigations} \\ \hline
Confidentiality & \begin{tabular}[c]{@{}c@{}}Eavesdropping\\ Data link interception\\ Man-in-the-middle\\ Identity spoofing\\ Hijacking\end{tabular} & Data link encryption \\ \hline
Integrity & \begin{tabular}[c]{@{}c@{}}Packet injection\\ Man-in-the-middle\\ Fabrication\\ Message deletion\\ Message modification\\ Replay attack\end{tabular} & \begin{tabular}[c]{@{}c@{}}Hash \\ Authentication\\ MAC\end{tabular} \\ \hline
Availability & \begin{tabular}[c]{@{}c@{}}Command and control data link spoofing\\ Channel jamming\\ Routing attack\\ Denial of service\\ Flooding\\ Buffer overflow\end{tabular} & Authentication \\ \hline
\end{tabular}

\end{table*}

\section{MAVSec: Security of the MAVLink protocol }
In this section, we propose MAVSec, a MAVLink enhanced version with cryptographic mechanisms to mitigate the vulnerabilities presented in the MAVLink protocol in terms of confidentiality. 

Cryptographic algorithms are classified as symmetric algorithms, which use symmetric keys or asymmetric algorithms, which use a couple of public/private keys.
Symmetric encryption is fast by design and consumes little energy, because the same key is used for both encryption and decryption. This makes it suitable for low-resource drones. However, asymmetric mechanisms can cause severe computational, memory, and energy overhead. Asymmetric cryptography is not suitable for static communications \cite{zhang2017improved}. 
 
Symmetric encryption algorithms are further classified in two basic categories: stream ciphers (such as RC4, CTR mode and ChaCh20) and block ciphers (such as AES). A block cipher encrypts fixed-length groups of N bits, called block of plaintext to a block of N bits of encrypted data, whilst a stream cipher can encrypt plaintext of varying sizes.

\textcolor{black}{ChaCha20 is a stream cipher developed by D. J. Bernstein in 2008, based on the Salsa cipher principles, to provide better diffusion and resistance against cryptanalytic attacks \cite{mouha2013proof}, without losing performance on software platforms \cite{bernstein2008chacha}. ChaCha20 is classified as a high-speed stream cipher although it is technically a block cipher in counter mode. For instance, ChaCha20 is often used by world leading companies like Google and Mozilla as it offers safer and faster alternatives \cite{bursztein2014speeding}.}

Advanced Encryption Standard (AES) is the most widely used symmetric cryptographic algorithm, which was chosen as a secure encryption algorithm by the National Institute of Science and Technology (NIST) \cite{standard2001announcing} among other encryption algorithms. AES is fast, flexible in block ciphers and has a high security and performance as compared to other symmetric encryption algorithms \cite{fips2011197}.

AES receives as input a plaintext block size of 128-bit and the encryption key length, which is either 128, 192 or 256 bits. The input text is processed by using the given key and applying a number of transformations to produce the output text (ciphertext).

AES block cipher algorithm can operate in five modes: Output Feedback (OFB), Electronic Code Book (ECB), Cipher Feedback (CFB), Cipher Block Chaining (CBC), and Counter (CTR) \cite{zhang2011comparative,lee2010price}. In this paper, we choose the CBC and CTR modes, since they are well known and widely used in encryption to encrypt MAVLink payload.

\subsection{Background on the encryption mechanisms.  }
In what follows, we describe the four cryptographic algorithms used in our experiments, namely, (1) the Advanced Encryption Standard in Counter Mode (AES-CTR), (2) the Advanced Encryption Standard in Cipher Block Chaining Mode (AES-CBC), (3) RC4 and (4)  ChaCha20.

\subsubsection{AES Counter Mode (CTR) }
The Counter mode (CTR) is a mode that turns a block cipher into a stream cipher and therefore used for achieving confidentiality \cite{lipmaa2000ctr}. First, a stream of input blocks is generated, called counters. The counters are obtained from an initial counter $IV$, which is incremented and used to encrypt each block in turn. Then a forward cipher function is applied to these counters to produce a sequence of output blocks $r_{i}$ that are exclusive-ORed with the plaintext $m_{i}$ to produce the ciphertext $c_{i}$.

Algorithm 1 explains the pseudo-code of AES-CTR encryption, where $IV$ is the initial counter value, $m_{i}$ represents the $ith$ block of plain text, and $c_{i}$ represents the $ith$ block of ciphertext. Both $IV + i$ and $m_{i}$ are independent.
Decryption transformation is identical to that of encryption. The main difference is that the plaintext and ciphertext positions are switched.

\begin{algorithm}[htb]
\DontPrintSemicolon
\SetAlgoLined
\SetKwInOut{Input}{Input}\SetKwInOut{Output}{Output}
\Input{$n$-block message $m= m_{1}\ldots m_{n}$, and a block cipher key $k$}
\Output{Ciphertext \textbf{AES-encrypt-ctr ($k,m$)} = $c_{1}\ldots c_{n}$}
\BlankLine
Initialization:\
 \BlankLine
$IV\leftarrow \left \{0,1\right\}^{n};$\;
$r_{0}\leftarrow IV;$    \;
\BlankLine
Encryption:\
\BlankLine
\For{i from 1 to n}{    
    
    \BlankLine
    
           $r_{i}=(IV+i)_{k}$ \\
        }
                
        \BlankLine
        
\For{i from 1 to n}{    
    
    \BlankLine
    
           $c_{i}=m_{i} \oplus r_{i} $ \\
        }
                
        \BlankLine
   return\  $c_{1} \ldots c_{n} $                      
        
\caption{Pseudo-code of CTR encryption.}
\end{algorithm}

    
    
                
        
    
    
                


Counter mode (CTR) is employed for its simplicity and efficiency because there is no need for a decoding function, nor for padding, and it offers a large flexibility in the implementation. Besides its high level of security \cite{lipmaa2000comments}, it presents high speed as it can be executed in parallel. Indeed, both encryption and decryption can be achieved in parallel on multiple blocks of plain or cipher data, which enables us to achieve a maximum level of parallelism. Another alternative is that CTR transforms block cipher into stream cipher, which is strongly recommended for our implementation since the stream cipher is more appropriate as MAVLink allows limited buffering.

\subsubsection{AES Cipher Block Chaining Mode (CBC) }
The Cipher Block Chaining (CBC) \cite{frankel2003aes} is a block cipher mode of operation, known to be the most commonly used whenever large amounts of data need to be sent securely. CBC mode chains the previous ciphertext block with the current message block before the cipher function. This mode is efficient at disguising any pattern in the plaintext: the encryption of each block depends on all the previous blocks.

 
Algorithm 2 explains the pseudo-code of AES-CBC encryption. As shown in Algorithm 2, the CBC mode takes a secret key $k$ as input, an Initialization Vector $IV$, which is randomly chosen with a length equal to the block length $N$, and the plaintext message. The plaintext is divided into several blocks $P_{1} \ldots P_{N}$, and each block is $XOR-ed$ with the cipher data of the previous block before it is encrypted. The result of the $XOR$ operation is encrypted with the key $K$ to produce ciphertext $C_{1} \ldots C_{N}$.
 
Decryption is thus the reverse process, which involves decrypting the current ciphertext and then adding the previous ciphertext block to the result. The IV and the encrypted message are sent to the recipient, which will then process this data using AES-CBC under the same key to check the integrity of the message and recover the plaintext message.

\begin{algorithm}[htb]
\DontPrintSemicolon
\SetAlgoLined
\SetKwInOut{Input}{Input}\SetKwInOut{Output}{Output}
\Input{$N$-block message $P= P_{1}\ldots P_{N}$, and a block cipher key $k$}
\Output{Ciphertext \textbf{AES-encrypt-cbc ($k,P$)} = $C_{1}\ldots C_{N}$}
\BlankLine
Initialization:\
 \BlankLine
$IV\leftarrow \left \{0,1\right\}^{n};$\;
$C_{0}\leftarrow IV;$    \;
\BlankLine
Encryption:\
\BlankLine
\For{i from 1 to N}{    
    
    \BlankLine
    
           $C_{i}=(P_{i}\oplus C_{i-1})_{k}$  \\
        }
                
        \BlankLine

   return\  $C_{1} \ldots C_{N}$                      
        
\caption{Pseudo-code of CBC encryption.}
\end{algorithm}

    
    
                



\subsubsection{RC4 }
It is the most popular and widely accepted symmetric key stream cipher algorithm in network security \cite{xie2010improved}.

The encryption of a message in RC4 is achieved by generating a keystream to be $ XOR-ed$  with a stream of plaintext to produce a stream of ciphertext.  The pseudo code for RC4 is shown in Algorithm 3. It has two parts: the first is a Key Scheduling Algorithm (KSA) whereas the second is the Pseudo-Random Number Generation Algorithm (PRGA), that generates a pseudo-random output sequence.

The KSA accepts the sized key $k$ as input, that may range between 8 and 2048 bits in multiples of 8 bits. It starts with the identity permutation in $S$ and uses the key continually swapping values to produce a new unknown key-dependent permutation. Since the only action on $S$ is to swap two values, the fact that $S$ contains a permutation is always maintained.

The PRGA  works by continually shuffling the permutation stored in $S$ as time goes on, each time picking a different value from the $S$ permutation as output. One round of RC4 outputs an n bit word as keystream, which is further $XOR-ed$ with the plaintext to produce the ciphertext.

\begin{algorithm}[htb]
  \caption{Pseudo code for RC4 stream cipher.}
  \begin{multicols}{2}
    \begin{algorithmic}
    \DontPrintSemicolon
\SetAlgoLined
			      \scriptsize
		\STATE	\textbf{\textsl{KSA}}
			\BlankLine
 \STATE Initialization:
 \BlankLine
\For{i from 0 to 255}{    

    \BlankLine
    
    $ S\lbrack i \rbrack=i$;   
        }
     $ j=0$; 
      \BlankLine

  L= length of the key.
   \BlankLine

  N= length of the Substitution box or state.
   \BlankLine

  K= key randomly chosen.
\BlankLine
 \STATE Scrambling:
\BlankLine

\For{i from 0 to N-1}{    
    
    \BlankLine
    
                $ j=(j+S\lbrack i \rbrack+K \lbrack i \bmod L \rbrack)$; 				
            $swap (S \lbrack i \rbrack,S\lbrack j \rbrack)$; 				
				}

        \BlankLine

    \end{algorithmic}
    \columnbreak
    \begin{algorithmic}
		\DontPrintSemicolon
\SetAlgoLined
      \scriptsize
      \STATE \textbf{\textsl{PRGA}}
 \BlankLine
  \STATE Initialization:
 \BlankLine
   $ i=0$;  
           \BlankLine

	 $ j=0$;     

\BlankLine
  \STATE Generation Loop:
\BlankLine    
    
          $ i=i+1$; 
                  \BlankLine

					$ j=j+S \lbrack i \rbrack$; 
					        \BlankLine

        $swap (S \lbrack i \rbrack,S\lbrack j \rbrack)$; 
                \BlankLine

				  $output O= S \lbrack S \lbrack i \rbrack + S\lbrack j \rbrack \rbrack$;

        \BlankLine
    \end{algorithmic}
  \end{multicols}
\end{algorithm}

The Stream cipher RC4 is efficient for real time processing. The algorithm is simple, fast and easy to explain. It can be efficiently implemented in both software and hardware.
\subsubsection{ChaCha20 }
\textcolor{black}{The ChaCha20 encryption algorithm requires the following parameters: a 256-bit encryption key, a 96-bit nonce, and a 32-bit Initial Block Counter to encrypt an arbitrary-length plaintext \cite{nir2018chacha20}}.\textcolor{black}{ The output is an encrypted message of the same length.
ChaCha20 generates a keystream by applying the ChaCha20 block function to the $key$, $nonce$, and a $block counter$. Plaintext is then encrypted using this keystream, with block $i$ of the plaintext $ XOR-ed$ with the output of the ChaCha20 block function, evaluated using the block counter $i$.}
\textcolor{black}{As the ChaCha20 block function is not applied directly to the plaintext, no padding should be necessary.}

\textcolor{black}{Decryption is performed in the same way. The ChaCha20 block function is used to expand the key into a keystream, which is $ XOR-ed$ with the ciphertext giving back the plaintext.}

\subsection{MAVSec: Integration of encryption mechanisms into MAVLink}
In this section, we introduce our proposed approach MAVSec and describe how the encryption mechanisms were implemented into the MAVLink protocol. 

The implementation involves the development and integration of the above-mentioned encryption algorithms both on the UAV autopilot side and GCS side integrated with the MAVLink protocol.

MAVLink messages contain a header with a MAVLink Identifier ID that cannot be encrypted. Therefore, only the MAVLink message payload can be encrypted since encrypting the header would result in the recipient being unable to recognize the appropriate MAVLink message type.

MAVLink makes use of a checksum to determine if a message was changed and therefore, the checksum needs to be recomputed after encryption. A solution to this would be to perform the encryption before calculating the checksum and decrypt after it is checked again.
 
Before sending any parameter through the payload, a heartbeat message is sent from the UAV to the GCS in order to verify that the system is ready and alive. The encryption is performed from the UAV to GCS. The payload is encrypted with the session key derived during the authentication phase, and the checksum is computed after encryption to ensure that the message is properly received by the ground station. The UAV sends a message containing the encrypted payload to the ground station. Once the message is received, the GCS, first checks the checksum and then decrypts the payload.  

The encryption algorithms AES-CTR, AES-CBC, ChaCha20 and RC4 are developed both on the autopilot and the GCS. The MAVLink source code is modified to include cryptographic functions resulting in a successful encryption and decryption. The payload is fed to the encryption algorithm as an input to obtain encrypted data. At the receiver's side, the received encrypted data is decrypted. The MAVSec code is available at Github \cite{mavsec}.  

\section{Experimental Validation}
In this section, we present a comprehensive study on the performance of the encryption algorithms integrated to the MAVLink protocol, in terms of the resource utilization such as CPU processing time and memory consumption rate. Based on the analysis of the obtained results, we discuss which algorithm is better to use according to the mentioned performance metrics.
\subsection{Simulation environment settings}

The experimental testbed consists of using a simulated drone with the Ardupilot Software-In-The-Loop (SITL)\cite{SITL}, which uses the same autopilot used in the real drone and the same MAVLink communication protocol. The use of a simulated drone generalizes straightforwardly to the case of a real drone. 
More specifically, the setup used for our experiments is as follows:
We used a virtual drone executed with the Ardupilot Software-In-The-Loop (SITL) simulator \cite{SITL}. It enables us to operate a Plane, Copter or Rover, without the need for any hardware. We have compiled the source code of Ardupilot to be able to integrate the encryption mechanisms into the message stream exchanged between the autopilot in the drone and the ground station. Besides, we used the QGroundControl \cite{qgroundcontrol} ground station, which is an open source ground control station (GCS) software application developed by Lorenz Meier and written in C++. To allow the secure communication between the autopilot of the SITL drone and the QGroundControl, we have also integrated the encryption algorithms into the QGroundControl to be able to decode the cipher stream received and extract the original MAVLink message. The GCS is connected to the simulated UAV via an open source GCS software called MAV proxy\cite{MAVProxy}.

The SITL simulator runs on a Linux virtual machine (Ubuntu 14.04 TLS) running on computer with 2.9 GHz Intel Core(TM) i7 CPU, 5.4 GB of memory. We used the ArduPilot version 3, more specifically, a UAV copter as a SITL to testing. To connect to the SITL, we used the port 14550 over the UDP protocol.

\subsection{Performance evaluation}
The experiments were performed ten times to make sure that the results comparing the different algorithms, are accurate and valid. Table \ref{tab:settings} shows the algorithm's parameters used in this experiment.
\begin{table}[htb]
\centering
\caption{\label{tab:settings} Algorithms settings.}
\begin{tabular}{|c|c|l|}
\hline
\textbf{Algorithm} & \textbf{\begin{tabular}[c]{@{}c@{}}Key size\\ \\ (Bits)\end{tabular}} & \multicolumn{1}{c|}{\textbf{\begin{tabular}[c]{@{}c@{}}Block size\\ \\ (Bytes)\end{tabular}}}             \\ \hline
AES-CTR            & 256                                                          & \begin{tabular}[c]{@{}l@{}}Any length, in our case is the \\ same length of the payload data\end{tabular} \\ \hline
AES-CBC            & 256                                                          & \multicolumn{1}{c|}{128}                                                                                  \\ \hline
RC4                & 256                                                           & \begin{tabular}[c]{@{}l@{}}Any length, in our case is the\\ same length of the payload data\end{tabular}  \\ \hline
ChaCha20            & 256                                                          & \begin{tabular}[c]{@{}l@{}}Any length, in our case is the \\ same length of the payload data\end{tabular} \\ \hline

\end{tabular}
\end{table}
Each algorithm was executed one after the other so that each one can have full system resources at its disposal. 

\begin{figure}[htb!]
	\centering
	\includegraphics[width=0.5\textwidth]{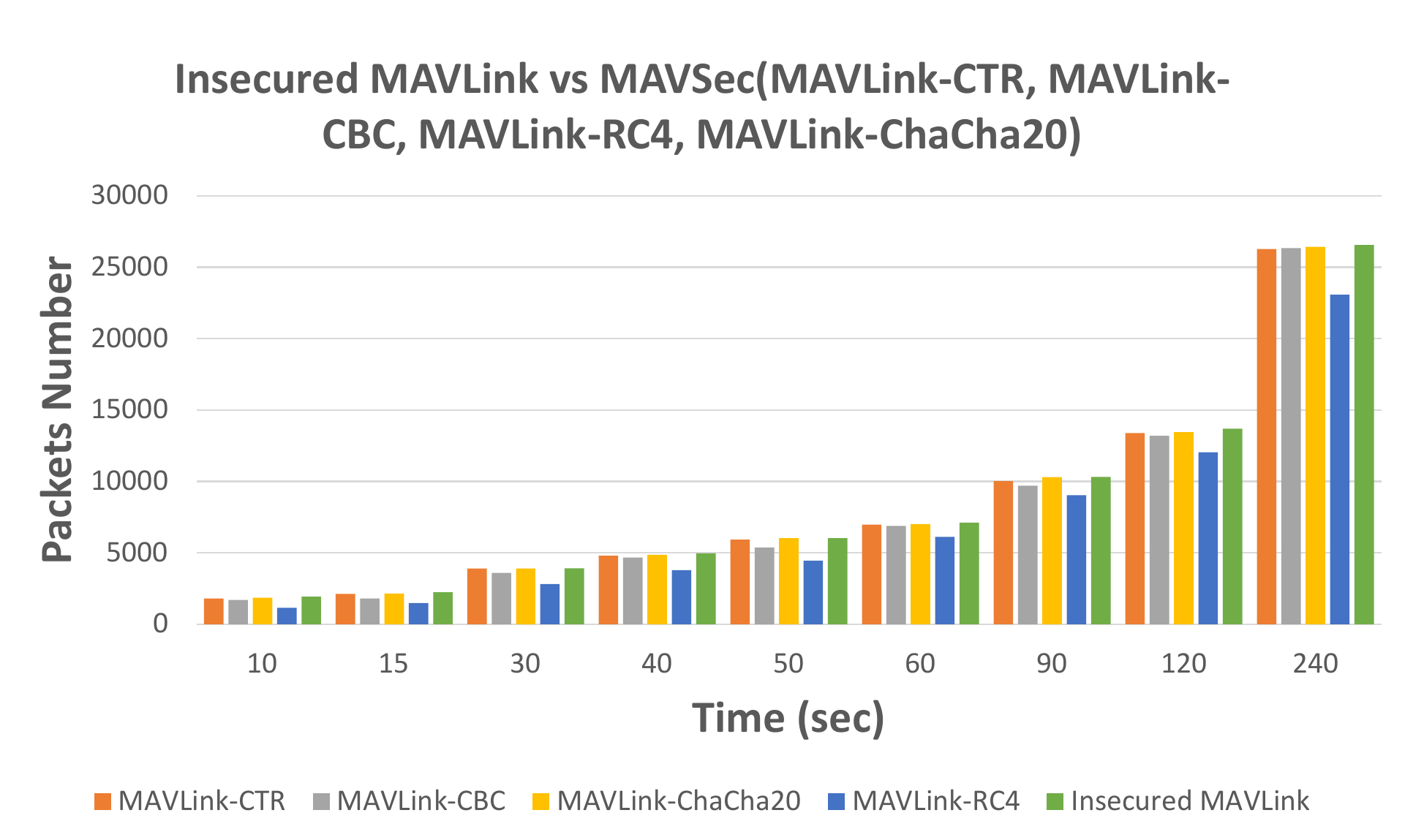}
	\caption{Transmitted packets number (insecured MAVLink vs MAVSec).}
	\label{fig.3}
\end{figure}

\textcolor{black}{To understand the effect of an encryption oriented solution, we compare the measured memory utilization, CPU consumption and packets transmitted of the insecured MAVLink protocol with the secured MAVLink protocol using cryptographic implementations (MAVSec). The  performance evaluation comparison between the  insecured MAVLink protocol and MAVSec can measure  the success of the implemented encryption mechanism. } 
\textcolor{black}{In terms of packet transmission, it can be clearly inferred from Fig.\ref{fig.3}, that MAVLink based stream cipher ChaCha20 and MAVLink-CTR send more packets compared to the MAVLink based CBC and RC4.} 

\textcolor{black}{As expected, the AES-CBC requires more processing time, so the number of transmitted packets will decrease. This can be explained by the key-chaining nature of the CBC and the fact that encryption is not performed in parallel. Also, the plaintext sizes that are not a multiple of the block size need to be padded which make the CBC unsuitable for encrypting and sending more packets. The AES-CTR mode encrypts and sends more packets because of its cipher type nature. } 
\textcolor{black}{MAVLink-ChaCha20 sends more packets as compared to MAVLink based CBC and CTR, because it is the fastest data transmission algorithm compared to AES \cite{nir2018chacha20}. Thus, ChaCha20 is suited to be used in lower powered UAV devices and real time communication. }

\textcolor{black}{The insecured MAVLink protocol sends more packets than the MAVLink-ChaCha20, but the difference is generally negligible.}

\textcolor{black}{Other important performance parameters are the memory and CPU utilization. Figures 4 and 5 show graphs that compare the average memory and CPU consumption of the insecured MAVLink with a MAVLink based ChaCha20, CTR, CBC and RC4.}

\textcolor{black}{As per the graph shown in figure 4, the MAVLink-RC4 is the most resource intensive, in terms of memory utilization, because the KSA and PRGA are executed sequentially in the RC4 encryption algorithm, which requires the use of more registers. However, the MAVLink-ChaCha20 takes less memory space, making ChaCha20 a good fit for UAV devices.
According to Fig. 4, no difference can be observed between the unsecured MAVLink protocol and the MAVLink-ChaCha20 in terms of memory usage.}

\textcolor{black}{The CPU usage is the percentage of time a CPU is committed for only a particular process of calculations. It reflects the load of the CPU. The more the CPU is used in the encryption process, the higher the load of the CPU will be \cite{salama2011studying}.} 
\textcolor{black}{The simulation results depicted in fig. 5 conclude that, MAVLink-RC4 takes the highest CPU utilization time period, wheras MAVLink-ChaCha20 consumes less CPU. This can be explained by the fact that ChaCha20 is based on ARX (Addition-Rotation-XOR) which are CPU friendly instructions \cite{redvzovic2016performance}. In contrast, the AES uses binary fields for the Sbox and Mixcolumns computations, which are generally implemented as a look up table to make it more efficient. 
Therefore, including ChaCha20 encryption algorithm will not affect the performance of the MAVLink protocol since MAVLink-ChaCha20 is very close in CPU utilization to the insecured MAVLink version.}

\begin{figure}[htb!]	\centering
	\includegraphics[width=0.5\textwidth]{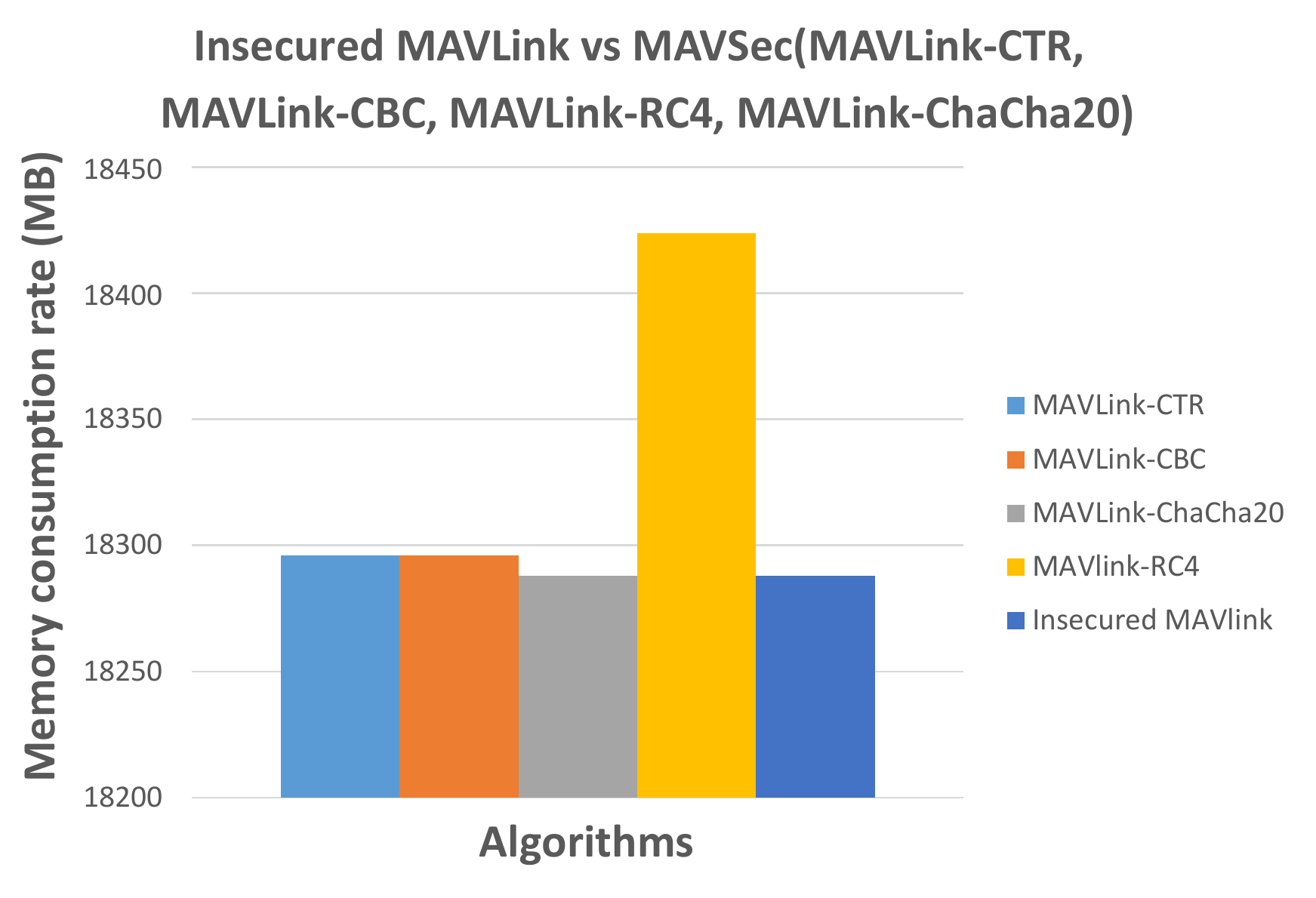}
	\caption{Memory consumption (insecured MAVLink vs MAVSec).}
	\label{}
\end{figure}

\begin{figure}[htb!]
	\centering
	\includegraphics[width=0.5\textwidth]{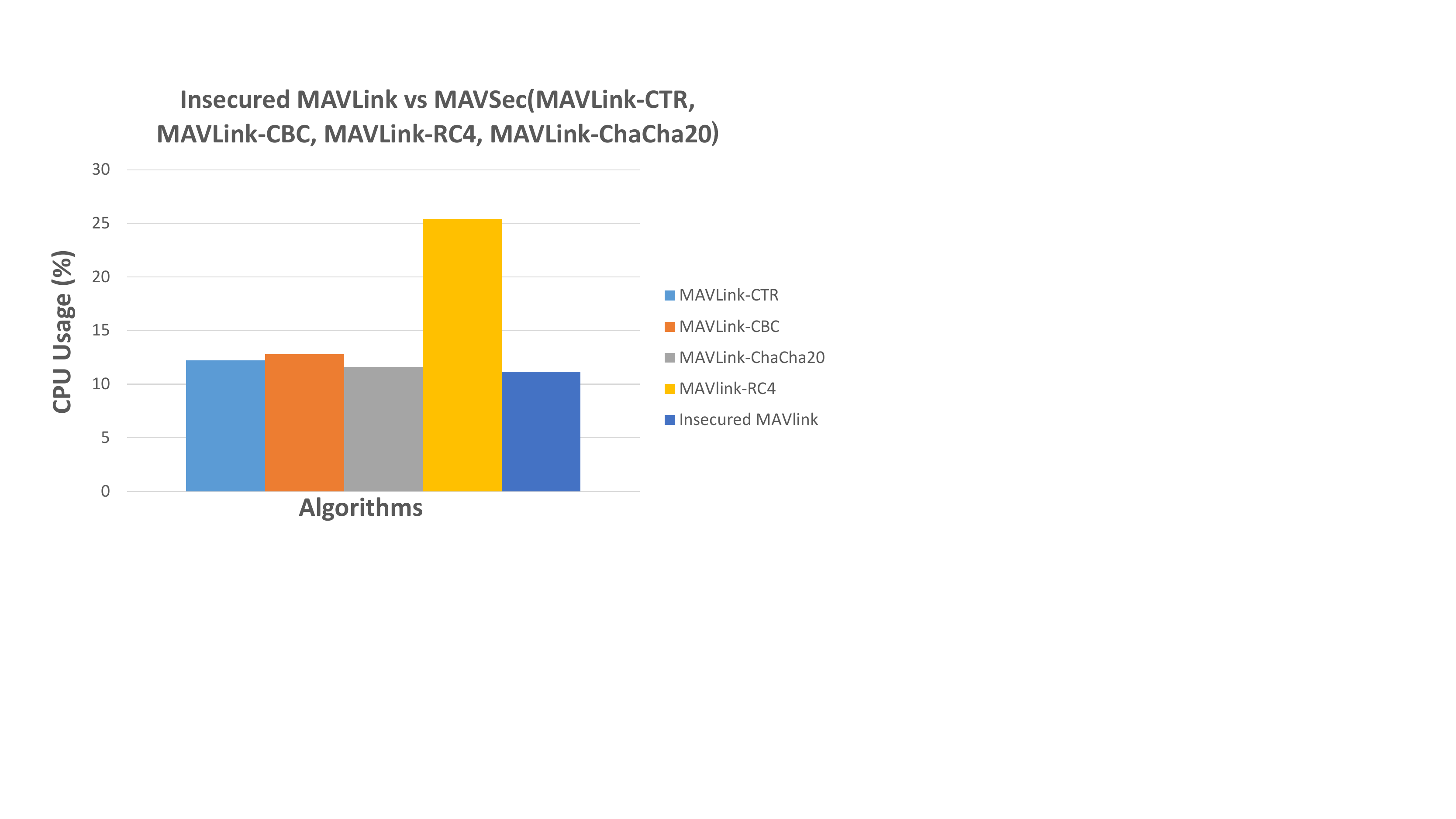}
	\caption{ CPU utilization (insecured MAVLink vs MAVSec).}
	\label{}
\end{figure}

\textcolor{black}{The performance evaluation comparison between the insecured MAVLink protocol and MAVSec allows us to measure the success of the implemented encryption mechanism.
Our set of simulation results were aimed to prove that ChaCha20 has better performance and is more efficient than other encryption algorithms. It can be considered as an excellent standard encryption algorithm to be adopted to secure MAVLink protocol and guarantee confidentiality of the MAVLink messages, without affecting its performance, consuming less memory space and CPU in order to preserve the memory and save the battery for resource-constrained drones.}


\section{Conclusion}
In this paper, we discussed the vulnerability and security threats against the MAVLink
protocol, then we proposed different cryptographic solutions to mitigate these vulnerabilities. The experimental study was achieved through implementing the encryption algorithms in the MAVLink source code. From the performance evaluation,\textcolor{black}{ we proved that ChaCha20 can be applied to secure the MAVLink protocol as it maintains confidentiality of the MAVLink messages without affecting its performance.}
In our future work, we will focus on testing and validating this implementation using a real scenario.

\section*{Acknowledgment}
This work is supported by the Robotics and Internet-of Things (RIOTU) Lab at Prince Sultan University.

\bibliographystyle{IEEEtran}
\bibliography{IEEEabrv,biblio}

\end{document}